\documentclass[12pt]{article}
\usepackage{amssymb,amsmath,amsthm,graphicx,ulem}
\usepackage{bm} 

\pdfoutput=1

\usepackage{graphicx,subfigure}
\usepackage{epsfig}
\usepackage{amsmath}
\usepackage{amsfonts}
\usepackage{amssymb}
\usepackage[usenames]{color}
\usepackage[a4paper,left=2.cm,right=2.cm,top=2.5cm,bottom=2.5cm]{geometry}
\usepackage{amsmath,amssymb,array,calc,rotating,epsfig,psfrag,cite}

\newcommand{\beq}{\begin{equation}}
\newcommand{\eeq}{\end{equation}}
\newcommand{\be}{\begin{equation}}
\newcommand{\ee}{\end{equation}}
\newcommand{\beqa}{\begin{eqnarray}}
\newcommand{\eeqa}{\end{eqnarray}}
\newcommand{\beqar}{\begin{eqnarray*}}
\newcommand{\eeqar}{\end{eqnarray*}}
\newcommand{\bea}{\begin{eqnarray}}
\newcommand{\eea}{\end{eqnarray}}

\numberwithin{equation}{section}

\long\def\eqlabel#1{\ifnum\draftcontrol=1
                    \tag@false  
                    \tag*{(\theequation) \hbox to -0.2cm{\hspace{0cm}\small{#1}\hss}}
                    \refstepcounter{equation}
                    \edef\@currentlabel{\theequation}
                    \ltx@label{#1}          
                    \else
                    \label{#1}
                    \fi
                    }

\def\revise#1       {\raisebox{-0em}{\rule{3pt}{1em}}%
                     \marginpar{\raisebox{.5em}{\vrule width3pt\
                     \vrule width0pt height 0pt depth0.5em
                     \hbox to 0cm{\hspace{0cm}{%
                     \parbox[t]{4em}{\raggedright\footnotesize{#1}}}\hss}}}}

\def\calf         {{\cal F}}

\def\calm         {{\cal M}}

\def\calo         {{\cal O}}

\def\zet          {{\mathbb Z}}

\def\del          {\partial}

\def\ee           {{\rm e}}

\def\sqr#1#2{{\vcenter{\vbox{\hrule height.#2pt  
 \hbox{\vrule width.#2pt height#1pt \kern#1pt
 \vrule width.#2pt}\hrule height.#2pt}}}}
\def\square{%
  \mathop{\mathchoice{\sqr{12}{15}}{\sqr{9}{12}}{\sqr{6.3}{9}}{\sqr{4.5}{9}}}}

\def\om{\Omega}

\def\ee{\epsilon}

\begin{document}




\title{Horizons cannot save the Landscape}

\vskip1cm

 \author{Iosif Bena${}^{\,a}$, Alex Buchel${}^{\,b,c}$, \'Oscar J. C. Dias${}^{\,a}$\\  \\ [0.2cm]
${}^{\,a}$ Institut de Physique Th\'eorique, CEA Saclay,\\[0.2cm]
 CNRS URA 2306, F-91191 Gif-sur-Yvette, France\\ [0.2cm]
 ${}^{\,b}$Department of Applied Mathematics\\
University of Western Ontario\\
 London, Ontario N6A 5B7, Canada\\[0.2cm]
${}^{\,c}$Perimeter Institute for Theoretical Physics\\
 Waterloo, Ontario N2J 2W9, Canada\\[0.2cm]
 \small{iosif.bena@cea.fr, abuchel@perimeterinstitute.ca, oscar.dias@cea.fr}}

 \date{}

 \maketitle 

\begin{abstract}

Solutions with anti-D3 branes in a Klebanov-Strassler geometry with
positive charge dissolved in fluxes have a certain singularity
corresponding to a diverging energy density of the RR and NS-NS three-form 
fluxes. There are many hopes and arguments for and against this
singularity, and we attempt to settle the issue by examining whether
this singularity can be cloaked by a regular event horizon. This is
equivalent to the existence of asymptotically Klebanov-Tseytlin or
Klebanov-Strassler black holes whose charge measured at the horizon
has the opposite sign to the asymptotic charge. We find that no such
KT solution exists. Furthermore, for a large class of KS black
holes we considered, the charge at the horizon must also have the same sign
as the asymptotic charge, and is completely determined by the
temperature, the number of fractional branes and the gaugino masses of the dual gauge theory.  
Our result suggests that antibrane singularities in
backgrounds with charge in the fluxes are unphysical, which in turn
raises the question as to whether antibranes can be used to
uplift AdS vacua to deSitter ones.  Our results also point out to a
possible instability mechanism for the antibranes. 

\bigskip
\bigskip
\bigskip
\bigskip

\noindent IPhT-T12/139\\UWO-TH-12/12

\end{abstract}

\newpage


\tableofcontents

\section{Introduction\label{sec:intro}}
The backreaction of antibranes in backgrounds with charge dissolved in
fluxes, like the Klebanov-Strassler (KS)\cite{ks} and the
Klebanov-Tseytlin (KT) solutions \cite{kt} has been a subject of
intense study over the past few years. A probe analysis of anti-D3
branes in the KS solution reveals that these branes have a nontrivial
potential that drives them to polarize into NS5 branes wrapping a
two-sphere inside the large three-sphere at the KS tip \cite{Kachru:2002gs}. 
On the other hand, if one tries to go beyond the probe approximation
and obtain the backreacted solution corresponding to smeared anti-D3
branes at tip of the KS solution, a surprise awaits: both the first-order
backreacted solution \cite{Bena:2009xk,Bena:2011hz,Massai:2012jn,Bena:2011wh}, as well as the fully backreacted
solution \cite{Bena:2012bk} have a singularity. 

The fate of this singularity is crucial. Adding anti-D3 branes to a
KS-like throat is the most generic way to uplift the vacuum energy of
the AdS vacua that come of out of string theory flux compactifications
with stabilized moduli \cite{Kachru:2003aw} and obtain deSitter
vacua. If the singularity of the anti-D3 solution is not physical,
this implies that anti-D3 branes cannot be used to give KS metastable
vacua and to uplift the AdS vacua to dS. Since the other known uplift
mechanisms (such as F or D-term uplifting \cite{other-uplift1, other-uplift2} or K\"ahler uplifting  
\cite{Balasubramanian:2005zx, other-uplift3}) are much
less generic, this would imply that string theory does not have a
landscape of dS vacua. Thus the fate of this singularity is closely
intertwined with that of the landscape.
  
In both the first-order and the fully backreacted solution, this singularity comes from three-form 
RR and NS-NS field strengths whose energy densities diverge. There have been quite a few arguments 
both in favor and against this singularity. The arguments in favor of this singularity \cite{Dymarsky:2011pm} are 
based on the self-consistency of the probe approximation of  \cite{Kachru:2002gs}  
and on the fact that the divergent energy of the singularity has a finite integral\footnote{In \cite{Dymarsky:2011pm} it was 
also argued that the singularity may be an artifact of first-order backreaction, but this has been shown not to happen 
\cite{Bena:2012bk}.}. The arguments against it are that the self-consistence of the probe approximation does not imply the existence of metastable 
vacua when backreaction is taken into account \cite{Bena:2006rg,Kutasov:2012rv}. Furthermore, the finiteness of the integral of the 
divergent energy density near a singularity can hardly constitute a criterion for accepting it: the negative-mass Schwarzschild 
solution also has a singularity with finite energy, and yet has to be discarded as unphysical 
\cite{Horowitz:1995ta}\footnote{Furthermore, there are similar singularities near anti-M2 \cite{Bena:2010gs,Massai:2011vi} 
and anti-D2 branes \cite{Giecold:2011gw}, and for those singularities both the energy density and its integral diverge.}.

It has also been argued that this singularity signals the tendency of the branes to polarize (as it happens in the probe 
approximation \cite{Kachru:2002gs}), and one can therefore hope that this singularity could be resolved by brane 
polarization\cite{Myers:1999ps} \`a la Polchinski-Strassler \cite{Polchinski:2000uf}. However, the recent calculation 
of \cite{Bena:2012vz}\footnote{As well as the earlier analysis of 
anti-D6 singularities \cite{Blaback:2010sj,Blaback:2011nz,Blaback:2011pn,Bena:2012tx}.} pours cold water on this hope: neither the smeared anti-D3 branes, nor the localized ones polarize into D5 branes wrapping the $S^2$ of the warped deformed conifold at a finite distance away from the KS tip despite the fact that {\it all} the terms of the Polchinski-Strassler polarization potential are there. Since in Polchinski-Strassler there are always multiple channels for brane polarization, the absence of a D5 brane polarization channel for localized branes suggests that the NS5 channel found in a brane probe approximation in  \cite{Kachru:2002gs} is also not present in the backreacted solution. 

Given that all the calculations made so far that could have either confirmed or invalidated the arguments in favor of this singularity have given negative results, the main hope that is left is that some hither unknown physical phenomenon will come to its rescue and resolve it. Even if this ``resolution by mystery'' proposal has not yet been articulated, it does appear to us that disproving it beforehand might once and for all settle the discussion.

The clearest argument that a given unknown singularity can be physical
has been formulated by Gubser in \cite{Gubser:2000nd}, who argued that
if the singularity can be cloaked by an event horizon it is physical,
and conversely if a singularity cannot be obtained by ``turning off''
a black hole horizon than it is not physical. Hence, if the anti-D3
singularity were physical, one would expect that there should
exist a black hole in Klebanov-Strassler and/or Klebanov-Tseytlin
whose charge has the opposite sign to the charge dissolved in the
fluxes. 

Our purpose is to revisit the KT and KS black holes
constructed numerically in the past and to show that no such black
hole exists. The existence
of Klebanov-Tseytlin black hole solution was first proposed in \cite{bh0} and then was 
constructed in perturbation theory around the black hole in
Klebanov-Witten \cite{Klebanov:1998hh} in
\cite{Buchel:2001gw,Gubser:2001ri}. However, the equations underlying
the nonlinear solution cannot be solved analytically, and the full
solution was constructed numerically in
\cite{aby05,Aharony:2007vg,Buchel:2009bh}\footnote{Other attempts to 
construct such black holes were discussed in \cite{pz1,pz2}.}.
 Interestingly-enough, extending the ansatz
to describe black holes in Klebanov-Strassler is not so hard, but it
turns out that these solutions (which would be dual to a deconfined
phase with spontaneously-broken chiral symmetry) do not exist
\cite{Buchel:2010wp}! The only way to build a black hole in
Klebanov-Strassler is to turn on one or two non-normalizable modes corresponding
to gaugino masses in the dual theory, which break explicitly the
chiral symmetry \cite{Buchel:2010wp}.

In this paper we review the one-parameter family of KT black holes,
and the three-parameter family of mass-deformed KS black holes and we
calculate the D3 charge at the horizon. We find that this charge does
not have an opposite sign to the asymptotic charge of the solution. In
fact, for a KT black hole of a given temperature the value of this
charge is not a free parameter, as one might expect naively from the
perturbative analysis of \cite{Gubser:2001ri}. Hence, if one imagines
keeping the temperature fixed and lowering probe charges into the
black hole, the configuration will settle back to the original charge.
This
comes essentially because the KT solution has charge dissolved in the
fluxes, and an over-charged or an under-charged black hole can expel
or absorb charge from the surrounding charge in the fluxes to bring
back its charge to its initial value. A similar story happens for the
mass-deformed KS black hole: if one adds positive or negative charge
to the black hole keeping the temperature and gaugino masses fixed,
the black hole will interact with the surrounding flux and return to
its original charge.

In section (\ref{sec:KTbh}) we discuss the KT black hole and examine
the symmetries of the background, the single-parameter nature of the
solution and the relationship between its temperature and the charge.
In section (\ref{sec:KSbh}) we explore in more detail the
three-parameter mass-deformed KS black hole of \cite{Buchel:2010wp} and explain the relation
between gaugino masses, temperature and charge. We conclude with a few
comments and suggestions for future work in section
(\ref{conclusion}).


\section{The Klebanov-Tseytlin black hole\label{sec:KTbh}}

\subsection{The cascading gauge theory and its  chiral symmetric phases\label{sec:KTphases}}

The   ${\cal N} = 1$ supersymmetric $SU(N)\times SU(N +M)$ cascading gauge theory can be realized as the world-volume theory on $N$ regular D3-branes to which we add $M$ fractional D3-branes (wrapped D5-branes) to the  apex of the conifold singularity. 
At low temperatures this cascading gauge theory spontaneously breaks a discrete chiral symmetry, which corresponds in the dual bulk to the deformation of the conifold singularity. The full warped deformed conifold solution dual to this cascading gauge theory was constructed by Klebanov and Strassler (KS) in \cite{ks}. One can also construct a singular solution dual to the chirally-symmetric phase of this theory (the  Klebanov-Tseytlin (KT) solution  \cite{kt}). The two solutions have D3 brane charges dissolved in the fluxes.

This cascading gauge theory is an ideal testing ground for
understanding the deconfinement phase transition of strongly-coupled
QCD-like gauge theories. The KS solution is holographically dual to
the confined chiral-symmetry-broken phase, and the high temperature
deconfined phase with unbroken chiral symmetry was argued in
\cite{bh0} to be dual to a black hole added to the KT solution. This
black hole was constructed in a perturbative expansion
\cite{Buchel:2001gw,Gubser:2001ri, aby05,Buchel:2009bh} in the fractional
brane number $M$ around the black hole in Klebanov-Witten
\cite{Klebanov:1998hh} and then at full nonlinear level in
\cite{Aharony:2007vg,Buchel:2009bh} using numerical methods.

We will review this solution with a highlight given to its properties that are relevant for our study: we emphasize that this is a one-parameter of solutions and we show that this black hole will always have positive Maxwell D3-brane charge at the horizon. Hence, there are no KT black hole solutions with anti-D3 brane charge, contrary to what one might have expected from the perturbative story.

\subsection{The Klebanov-Tseytlin black hole\label{sec:KTbhConstruction}}

We review the construction of the  Klebanov-Tseytlin black hole, which is a solution of 10-dimensional type IIB supergravity that asymptotes to the Klebanov-Tseytlin (KT) background \cite{kt}, and is holographically dual to a high-temperature deconfined chirally-symmetric phase of the cascading gauge theory. Recall that the IIB supergravity field content includes, besides the gravitational field, a dilaton $\Phi=\ln g$, a NS-NS flux $H_3=dB_2$, and R-R fluxes $F_3$ and $F_5$ (the axion vanishes, $C_0=0$). 
These solutions have a $Z_{2M}$ chiral symmetry, a $U(1)_B$ symmetry and a $SU(2)\times SU(2)$
global symmetry. In the absence of the black hole, the KT background is a direct product of  $\calm_5$ with metric $g_{\mu\nu}$ and a squashed $T^{1,1}$ with the radii of curvature of $\calm_5$ and the fluxes $H_3,\,F_5$  varying logarithmically in the radial coordinate.

The most general ansatz (in the Einstein-frame) that is a deformation of the KT background \cite{kt} which preserves its $SU(2)\times SU(2)\times \zet_{2P}\times U(1)_B$ symmetry, is \cite{aby05,Aharony:2007vg}
\begin{equation}
ds_{10}^2 =g_{\mu\nu}(y) dy^{\mu}dy^{\nu}+ds^2_{T^{1,1}}\,, \qquad ds^2_{T^{1,1}}=\om_1^2(y) e_{\psi}^2 
+\om_2^2(y) \sum_{a=1}^2\left(e_{\theta_a}^2+e_{\phi_a}^2\right),
\label{10met}
\end{equation}
and (we set $\alpha^\prime\equiv 1$)
\begin{equation}
\begin{split}
&F_3=P\ e_\psi \wedge \left(e_{\theta_1}\wedge e_{\phi_1}-e_{\theta_2}\wedge e_{\phi_2}\right)\,,\qquad
B_2=\frac{K(y)}{2 P}\ \left(e_{\theta_1}\wedge e_{\phi_1}-e_{\theta_2}\wedge e_{\phi_2}\right),\\
&F_5=\calf_5+\star \calf_5\,,\qquad \calf_5=-K(y)\ e_\psi\wedge e_{\theta_1}\wedge e_{\phi_1}\wedge e_{\theta_2}\wedge e_{\phi_2}\,,
\end{split}
\label{fluxforms}
\end{equation}
where $y$ denotes the coordinates of $\calm_5$ (greek indices $\mu,\nu$ run from $0$ to $4$) and $ds^2_{T^{1,1}}$ is the line element of the warp-squashed $T^{1,1}$, with the one-forms $e_{\psi},\ e_{\theta_a},\ e_{\phi_a}$ ($a=1,2$) given by:
\begin{equation}
e_{\psi}=\frac 13 \left(d\psi+\sum_{a=1}^2 \cos\theta_a\ 
d\phi_a\right),\qquad
e_{\theta_a}=\frac{1}{\sqrt{6}} d\theta_a\,,\qquad 
e_{\phi_a}=\frac{1}{\sqrt{6}} \sin\theta_a\ d\phi_a.
\label{1forms}
\end{equation}
The range of the $T^{1,1}$ coordinates is $0\leq \psi \leq 4\pi$, $0\leq \theta_a\leq \pi$ and $0\leq \phi_a\leq 2\pi$.
 The dimensionful constant $P$ in \eqref{fluxforms} is related to the quantized dimensionless units of flux $M$ entering in the rank of the gauge groups of the dual field theory. We work in the normalization where \begin{equation}\label{defPvsM}
P=\frac{3}{2^{3/4} \pi} G_5^{1/4} M\,,
\end{equation}
where $G_5=G_{10}/{\rm vol}_{T^{1,1}}$ is the 5-dimensional Newton's constant obtained after the dimensional reduction on $T^{1,1}$ done next.

With this ansatz, we can do a standard Kaluza-Klein reduction of the type IIB action to five dimensions and get the effective action \cite{aby05}
\begin{equation}
\begin{split}
S_5= \frac{1}{16\pi G_5} \int_{\calm_5} {\rm vol}_{\calm_5}\
 \biggl\lbrace &
\Omega_1 \Omega_2^4 \biggl(R_{10}-\frac 12 \nabla_\mu \Phi \nabla^\mu
\Phi\biggr)-P^2 \Omega_1 e^{-\Phi}\biggl(
\frac{\nabla_\mu K\nabla^\mu K}{4P^4}+\frac{e^{2\Phi}}{\Omega_1^2}
\biggr)\\
&-\frac 12\frac{K^2}{\Omega_1\Omega_2^4}\biggr\rbrace,
\end{split}
\label{5action}
\end{equation}
where $R_{10}$ is the 10-dimensional Ricci scalar related to the 5-dimensional Ricci scalar $R_5$ by
\begin{equation}
\begin{split}
R_{10}=R_5&-2\om_1^{-1}g^{\lambda\nu}\biggl(\nabla_{\lambda}\nabla_{\nu}\om_1
\biggr)-8\om_2^{-1}g^{\lambda\nu}\biggl(\nabla_{\lambda}\nabla_{\nu}\om_2
\biggr)\\
&-4 g^{\lambda\nu}\biggl(2\ \om_1^{-1}\om_2^{-1}\ 
\nabla_\lambda\om_1\nabla_\nu\om_2
+3\ \om_2^{-2}\ \nabla_\lambda\om_2\nabla_{\nu}\om_2\biggr)\\
&+24\ \om_2^{-2}-4\ \om_1^2\ \om_2^{-4},
\end{split}
\label{ric5}
\end{equation}
The associated equations of motion are \cite{aby05}
\begin{eqnarray}
&& 0=\frac{1}{\sqrt{-g}}\ \del_\mu\left[\frac{e^{-\Phi}\om_1}
{2P^2}\sqrt{-g} g^{\mu\nu}\del_\nu K\right]-\frac{K}{\om_1\om_2^4}, \nonumber \\
&&
0=\frac{1}{\sqrt{-g}}\del_\mu\left[\om_1\om_2^4\sqrt{-g}
g^{\mu\nu}\del_\nu\Phi\right]+\frac{\om_1e^{-\Phi}(\del K)^2}{4P^2}
-\frac{P^2e^\Phi}{\om_1},\nonumber \\
&&
0= \om_2^4 R_5-12\om_2^2(\del\om_2)^2+24\om_2^2-12\om_1^2-8\om_2^3\square_5\om_2 \nonumber \\
&& \hspace{0.8cm} 
-\frac 12
\om_2^4(\del\Phi)^2+\frac{P^2e^\Phi}{\om_1^2}-\frac{e^{-\Phi}(\del K)^2}
{4P^2}+\frac {K^2}{2\om_1^2\om_2^4},\nonumber \\
&&
0=\ 4\om_1\om_2^3R_5-8\om_2^3\square_5\om_1-24\om_1\om_2^2\square_5\om_2
-24\om_2^2\del\om_1\del\om_2-24\om_1\om_2(\del\om_2)^2+48\om_1\om_2\nonumber \\
&&\hspace{1cm}
-2\om_1\om_2^3(\del\Phi)^2+\frac{2K^2}{\om_1\om_2^5},\nonumber \\
&&
\om_1\om_2^4\ R_{5\mu\nu}=\frac{g_{\mu\nu}}{3}
\left\{\frac{P^2e^\Phi}{\om_1}+\frac{K^2}{2\om_1\om_2^4}
+\square_5\left(\om_1\om_2^4\right)-24\om_1\om_2^2+4\om_1^3\right\} \nonumber \\
&& \hspace{2.5cm} 
+\nabla_\mu\nabla_\nu\left(\om_1\om_2^4\right)-4\om_2^3
\left(\del_\mu\om_1\del_\nu\om_2+\del_\nu\om_1\del_\mu\om_2\right)
-12\om_1\om_2^2\del_\mu\om_2\del_\nu\om_2\nonumber \\
&&\hspace{2.5cm} 
+\frac{\om_1e^{-\Phi}}{4P^2}\ \del_\mu K\del_\nu K
+\frac 12 \om_1\om_2^4\ \del_\mu \Phi\del_\nu \Phi,   
\label{eom}
\end{eqnarray}
where $(\del F)^2$ denotes $g^{\mu \nu} \del_{\mu} F \del_{\nu} F$ and
$\square_{\,5}$ is the Laplacian in $\calm_5$.

An ansatz for the  $\calm_5$ gravitational field that is tailored to search for the KT black hole with horizon located at $x=1$ and asymptotic KT background at $x\to 0$ is \cite{Aharony:2007vg}
\begin{equation}
ds_{\calm_5}^2=h^{-1/2}(2x-x^2)^{-1/2}\left[-(1-x)^2 dt^2+dx_1^2+dx_2^2+dx_3^2\right]+G_{xx} dx^2,
\label{ktm}
\end{equation}
and it is also useful to rewrite the squashed $T^{1,1}$ factors as
\begin{equation}
\Omega _1=h^{1/4}\sqrt{f_2}\,,\qquad \Omega _2=h^{1/4}\sqrt{f_3}\,.
\label{def:fs}
\end{equation}
In these expressions, $h$, $f_2$, $f_3$, $G_{xx}$ are functions of the compact radial
coordinate $x$. Recall that the solution is fully specified once we find in addition the  dilaton  $\Phi=\ln g(x)$, and flux forms \eqref{fluxforms}, which are determined by a single function $K(x)$.

In these conditions, it follows from the equations of motion \eqref{eom} that $G_{xx}$ is given by an algebraic relation that is a function of $\{ h, \, f_2,\,f_3,\,K,\,g \}$ and their first derivatives. So, in total, to determine our black hole we need to solve a system of five second-order coupled and non-linear ODEs for $\{ h, \, f_2,\,f_3,\,K,\,g \}$. The explicit algebraic relation for $G_{xx}$ and the system of five equations of motion are explicitly written in equations (2.6)-(2.12) of \cite{Aharony:2007vg}.
An important observation is that each of these five equations is second order. Therefore the total differential order of the system is $10$.

To find the KT black hole we solve the equations of motion subject to appropriate boundary conditions at the horizon, $x\to 1$, and at the asymptotic boundary $x\to 0$.

The IR boundary condition (horizon) is quite standard in black hole physics. Formally, we rewrite our ansatz in ingoing Eddington-Finkelstein coordinates (appropriate to extend the analysis through the horizon) and we require regularity of all our fields at the horizon in this coordinate system. This amounts to say that all our functions must have a series expansion in $(1-x)^2$ \cite{Aharony:2007vg}, 
\begin{equation}
\begin{split}
h=&\sum_{n=0}^\infty h_{n}^h\ (1-x)^{2n},\qquad f_2=\sum_{n=0}^\infty
a_{n}^h\ (1-x)^{2n}\,,\qquad
f_3=\sum_{n=0}^\infty b_n^h\ (1-x)^{2n}\,,\\
g=&\sum_{n=0}^\infty g_n^h\ (1-x)^{2n}\,, \qquad K=\sum_{n=0}^\infty k_n^h\ (1-x)^{2n}\,. \\
\end{split}
\label{seriesH}
\end{equation}
with the boundary condition simply requiring that the coefficients $\{ h_{0}^h,\,a_{0}^h,\,b_{0}^h,\,k_{0}^h,\,g_{0}^h \}$ are constants. Solving the five equations of motion perturbatively around the horizon up to arbitrary order one finds that  only a few of the coefficients in \eqref{seriesH} are independent, with all the others being a function of these. Concretely, one chooses the 6 IR independent parameters to be
\begin{equation}
\hbox{IR independent parameters (6):}\qquad  \{h_0^h,\, a_0^h,\,b_0^h,\,k_0^h,\,g_0^h,\,a_1^h\}.
\label{paramH}
\end{equation}

Consider now the UV asymptotic structure at $x\to 0$. The UV boundary condition is straightforward: we want the KT black hole to approach asymptotically the KT solution \cite{kt}. The latter is the leading order contribution of a power series expansion in $x$ and $\ln(x)$ of the equations of motion \cite{Aharony:2007vg},
\begin{equation}
\begin{split}
h=& h_0- \frac{P^2 g_0}{8a_0^2}\ \ln (x)+\sum_{n=1}^{\infty}\sum_{k=1}^{n-1} h_{n,k}\ x^{n/2}\ \ln^k (x), \qquad
f_2=a_0+\sum_{n=1}^{\infty}\sum_{k=1}^{n-1} a_{n,k}\ x^{n/2}\ \ln^k (x), \\
f_3=& a_0+\sum_{n=1}^{\infty}\sum_{k=1}^{n-1} b_{n,k}\ x^{n/2}\ \ln^k (x), \qquad g= g_0+\sum_{n=1}^{\infty}\sum_{k=1}^{n-1} g_{n,k}\ x^{n/2}\ \ln^k
(x) \\
K=& 4 h_0 a_0^2-\frac 12 P^2 g_0-\frac 12 P^2 g_0 \ln (x)+\sum_{n=1}^{\infty}\sum_{k=1}^{n-1} K_{n,k}\ x^{n/2}\ \ln^k (x).
\end{split}
\label{seriesBdry}
\end{equation}
The presence of the $\ln x$ terms in $h, K$ in the KT leading contribution is responsible for the logarithmic running of the fluxes and  $\calm_5$ radii of curvature. 

As expected, not all coefficients in the expansion \eqref{seriesBdry} are independent. There are 4 independent asymptotic parameters already present in the KT solution and that we choose to be $\{P,\,g_0,\,a_0,\,h_0\}$: $P$ is related to the quantized flux $M$ by \eqref{defPvsM},  $g_0$ is related to the
dimensionless parameter of the cascading gauge theory, and there are two combinations of the other 2 parameters that are related to the temperature and dynamical scale of the cascading gauge theory. 
More concretely, one has $a_0^2=4\pi G_5\ sT$, where $s$ and $T$ are the entropy density  and temperature of the solution, which can be computed directly from the horizon area and surface gravity using the expansion \eqref{seriesH}. In addition we can replace the independent parameter $h_0$ by a new dimensionless parameter $k_s$ defined as,
\begin{equation} \label{def:ks}
k_s\equiv \frac{4 h_0 a_0^2}{P^2 g_0}-\frac 12 \equiv \frac{1}{2} \ln\left(\frac{4 \pi G_5 s T}{\Lambda^4}\right) \,.
\end{equation}
This relation defines  the dynamical scale $\Lambda$ of the cascading theory.
The other UV independent coefficients correspond to vacuum expectation values (VEVs) of the operators dual to the fields we are solving for. After using the conformal anomaly equation  \cite{aby05}, one finds that there are four VEV independent parameters, chosen to be $\{a_{2,0},\, g_{2,0},\,a_{3,0},\,a_{4,0}\}$.  

We conclude that there are eight UV independent parameters, that determine the coefficients in \eqref{seriesBdry} to any order we wish.  These can be chosen to be $\{P,\,g_0,\,a_0,\,k_s,\, a_{2,0},\, g_{2,0},\,a_{3,0},\,a_{4,0}\}$.
We can however use the symmetries of the problem to get rid of some of these parameters.
First observe that that the ansatz \eqref{ktm} is invariant under the scaling symmetry:\footnote{This scaling symmetry leaves $h_0 a_0^2$ invariant. This means that  $h_0 a_0^2$  is a function of the dimensionless parameter of our theory, which is the ratio between the temperature
and the dynamical scale $\Lambda$. This motivates the definition \eqref{def:ks}.}
\begin{equation}
 (t,x,\vec{x}) \to (\lambda^{-2} t,x, \lambda^{-2}\vec{x})\,,\qquad
h\to \lambda^{-2}\ h\,,\qquad f_{2,3}\to \lambda f_{2,3}\,,\qquad  g\to g\, \qquad  K\to K\,.
\label{scaling}
\end{equation}
We use this scaling symmetry to set $a_0\equiv 1$.
In addition, we are working in the supergravity approximation (where we neglect  $g_s$ and $\alpha'$ corrections). In this approximation,  the action and the associated equations of
motion do not depend separately on $P^2$ and $g_0$ but only on the combination $P^2 g_0$. We can thus set $g_0\equiv 1$.
Furthermore, when we neglect $\alpha'$ corrections, the action is
multiplied by a constant when we rescale the ten dimensional metric
by a constant factor (and rescale the $p$-forms accordingly), so
that the equations of motion are left invariant. This transformation
acts on our variables as
\begin{equation} \label{scaling2}
h\to \lambda^{-2} h\,,\qquad f_{2,3} \to \lambda^2 f_{2,3}\,,\qquad K\to
\lambda^2 K\,,\qquad g\to g\,, \qquad P\to \lambda P\,.
\end{equation}
As long as we work in the supergravity approximation we can thus use this symmetry to set $P\equiv 1$ and use the scaling \eqref{scaling2}  to generate the solutions for any other value of $P$.

To sum, after using the symmetries of our system we find that we are left with 5 independent UV parameters,
\begin{equation}
\hbox{UV independent parameters (5):}\qquad  \{\,k_s,\, a_{2,0},\, g_{2,0},\,a_{3,0},\,a_{4,0}\}.
\label{paramBdry}
\end{equation}

At this point we can ask how many parameters we need to describe the KT black hole solution. We have a total of $6+5=11$ IR and UV independent parameters \eqref{paramH} and \eqref{paramBdry}, and the equations of motion are a system of total differential order 10. We thus conclude that the KT black hole is a one-parameter family of solutions.  In the numerical search of the KT black hole we will take this dimensionless parameter to be $k_s$. Equation \eqref{def:ks} can then be used to translate the results we obtain in terms of the dimensionless temperature $T/\Lambda$.

The black hole is constructed numerically using a standard shooting method.
One fixes the microscopic free parameter of the theory, $k_s$, and wants to find the other ten IR and UV independent parameters \eqref{paramH} and \eqref{paramBdry}.  The equations of motion have two critical points: at the horizon, $x=1$, and at the 
boundary, $x=0$. Consider first the horizon. Using the IR series expansion \eqref{seriesH}, one constructs the solution in the near-horizon region up to the tenth order  ($n=5$ inclusive) in the radial distance to
the horizon. This solution depends on the 6 IR parameters \eqref{paramH}. One then integrates
numerically the six radial second order ODEs, using a Runge-Kutta method, up to a large radial distance. 
The procedure is repeated, this time at the asymptotic critical point where one
starts by obtaining the asymptotic solution up to ninth order ($n=9$, inclusive) using the UV series expansion \eqref{seriesBdry}. Fixing $k_s$ this asymptotic solution  is a function of the four UV parameters $\{a_{2,0},\, g_{2,0},\,a_{3,0},\,a_{4,0}\}$. One integrates this solution down to very small
values of the radial distance. In the overlapping region of the two
solutions we then do their matching, more precisely at $x=0.5$. The requirement that both the set of five functions  $\{ h, \, f_2,\,f_3,\,K,\,g \}$ and their first derivatives must be continuous gives ten conditions that we use to fix 
the 10 IR and UV independent parameters (for a given $k_s$).
The whole process is now repeated for other values of $k_s$ to generate the results that will be presented in the next subsection.

\subsection{Properties of the Klebanov-Tseytlin black hole\label{sec:KTbhProperties}}

We are ready to discuss some physical properties of the Klebanov-Tseytlin black hole that are relevant for our study. For a  detailed account of other properties see \cite{Aharony:2007vg,Buchel:2009bh}.

We present the results as a function of either the microscopic parameter $k_s$ (which defines uniquely the solution) or as a function of the dimensionless quantity $T/\Lambda$, where $T$ is the temperature of the solution and $\Lambda$ is the dynamical scale of the theory, related to $k_s$ via \eqref{def:ks}. As explained above, the numerical construction of the solution exploits the scaling symmetries of the theory, and is done at $P=g_0=a_0\equiv 1$. One can then restore the correct powers of $P$ using the scaling symmetry \eqref{scaling2}, which also introduces a factor of $g_0/a_0$ together with every factor of $P^2$. In order to relax the $a_0=1$ condition we then use
\eqref{def:ks} which implies that all the dimensionful quantities must be computed in units of  $\Lambda=e^{-k_s/2}$.  

From the surface gravity we can compute the black hole temperature, which in $\Lambda$ units is
\begin{equation}
\frac{T}{\Lambda}=\frac{e^{k_s/2}}{4\pi h_0^h b_0^h}\ \sqrt{\frac{2
(8 h_0^h (a_0^h)^2-g_0^h P^2)}{a_0^h+2 a_1^h}}\,, \label{sTl}
\end{equation}
and enables an immediate translation of a dependence on $k_s$ into a dependence
on the temperature. The entropy density is computed from the horizon area.
The energy density ${\cal E}$ and pressure ${\cal P}$ of the KT deconfined plasma phase can be read from the expectation values of the appropriate components of the holographic stress-tensor $T_{ab}$: ${\cal E}=\langle T_{tt}\rangle$ and ${\cal P}=\langle T_{x_i x_i}\rangle$. Moreover, since there are no chemical potentials, the free energy density is a simply ${\cal F}=-{\cal P}$. These quantities are dimensionless if expressed in $\Lambda$ units, and given by 
\begin{equation}
\frac{4\pi  G_5}{P^2g_0^2}\frac{s}{T^3}= \left(\frac{T}{\Lambda}\right)^{-4}\! e^{2 k_s} ,\quad
\frac{32\pi^4}{81}\frac{\cal E}{\Lambda^4} =\frac 34\left(1+\frac{4}{7}
\frac{a_{2,0}}{a_0}\right)e^{2 k_s},
\quad \quad\frac{32\pi^4}{81}\frac{\cal F}{\Lambda^4} =\frac 37\left(\frac{a_{2,0}}{a_0}-\frac{7}{12}\right)e^{2 k_s}. \label{efl}
\end{equation}

\begin{figure}[th]
\centering
\includegraphics[width=.47\textwidth]{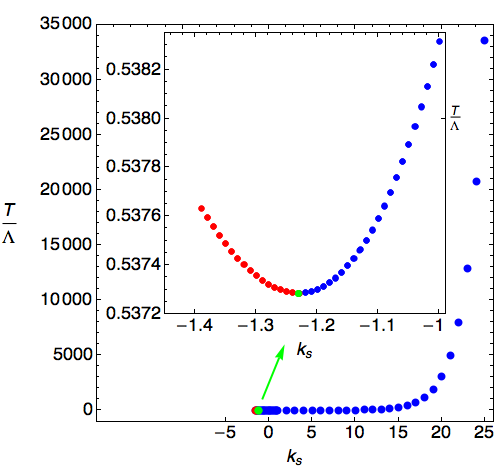}
\hspace{0.5cm}
\includegraphics[width=.47\textwidth]{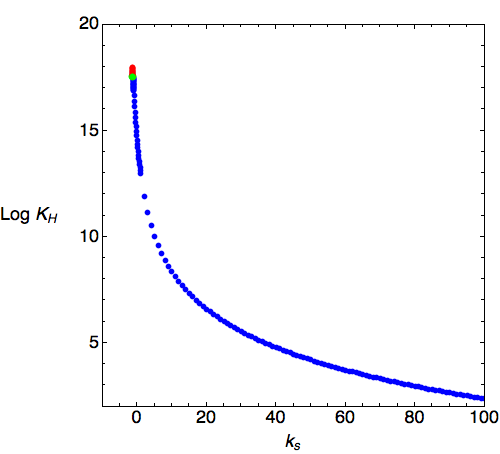}
  \caption{
{\bf Left Panel:} The dimensionless temperature $\frac{T}{\Lambda}$ as a function of the microscopic parameter $k_s$. The inset plot zooms it in the neighbourhood of $T=T_u$ (green point). The red dots  correspond to $k_s\le k_{u}$ and the blue dots correspond to $k_s\ge k_{u}$. To the left of the last red dot there are no black holes.
{\bf Right Panel:} The Kretschmann scalar at the horizon as a function of the microscopic parameter $k_s$. } \label{figure1}
\end{figure}

We can now describe the KT black hole using these thermodynamic
quantities.  The microscopic parameter $k_s$ has no positive upper
bound. For large positive values of $k_s$, the temperature is large
and then decreases as $k_s$ decreases (see {\it Left panel} of
Fig. \ref{figure1}). However, as we keep decreasing $k_s$, the
temperature reaches a minimum at $k_s=k_s^{u}=-1.230(3)$ (green dot in
Fig. \ref{figure1}) and, for even smaller $k_s$, it increases until a
critical lower bound, $k_s=k_s^{crit}$, is reached (represented by the
last numerical red point in Fig. \ref{figure1}). As this critical
value is approached the curvature invariant (like the Kretschmann
scalar in the Einstein frame) grows without bound, as shown in the {\it Right panel} of
Fig. \ref{figure1}, which implies that the solution will develop a naked singularity. 
Below $k_s=k_s^{crit}$ there are no longer KT black
holes. In Fig. \ref{figure1}, the inset plot shows the behavior of the
temperature around its minimum value.  In all the plots of this
Section, the numerical red dots describe the region with $k_s\le
k_{u}$ and the blue dots correspond to $k_s\ge k_{u}$.

\begin{figure}[th]
\centering
\includegraphics[width=.462\textwidth]{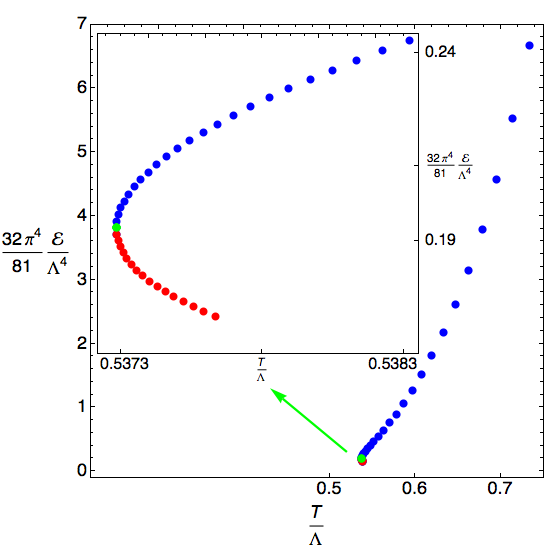}
\hspace{0.5cm}
\includegraphics[width=.488\textwidth]{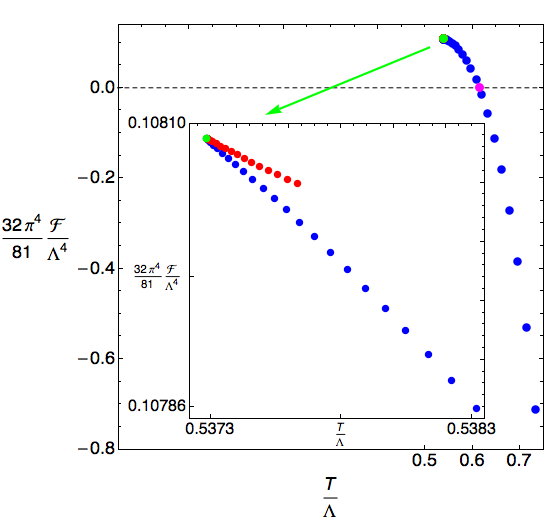}
  \caption{
  {\bf Left Panel:} The dimensionless energy density as a function of the dimensionless temperature. The inset plot shows this energy in the vicinity of the onset of the perturbative instability of the cascading plasma (where $T=T_u$, the green point). 
  {\bf Right Panel:}  The dimensionless free energy density as a function of the temperature for temperatures at the deconfinement transition ($T=T_c$; magenta point) and around it. The inset plot shows the  free energy density in the vicinity of the perturbative instability of the cascading plasma ($T=T_u$; green point) and all the way down to the last red point that represents the last KT black hole that exists. 
 In these figures, the red/blue dots have the same meaning as in Fig. \ref{figure1}. 
} \label{figure2}
\end{figure}

The {\it Left panel} of  Fig. \ref{figure2} shows what happens to the energy density as a function of the temperature in the most interesting region of the parameter space. For larger temperatures it grows monotonically. On the other hand, the free energy density as a function of the temperature is plotted in the {\it Right panel} of Fig.  \ref{figure2}.
For large values of the temperature the KT black hole solution is the preferred thermodynamic phase since its free energy is negative. However, it vanishes at \footnote{To pinpoint with higher accuracy the critical temperatures  of the green ($T=T_u$) and magenta points  ($T=T_c$) we have made runs in the code in a small window around these points with much higher resolution than the one presented in the figures.}  
\begin{equation}\label{confT}
T_c=0.6141111(3)\, \Lambda\,,
\end{equation}
and is positive for smaller temperatures: a first order deconfinement/confinement phase transition occurs at $T=T_c$ (represented by the magenta point in Fig.  \ref{figure2}). 
The inset plot in this right panel shows how the free energy behaves in the vicinity of $T=T_u$ (green point). We see that in this region the free energy keeps positive but precisely at $T=T_u\left(k_s^{u}\right)$ with
 \begin{equation}
T_u=0.8749(0)\, T_c
\end{equation}
 we find a cusp, and a continuous phase transition occurs. To the right of this cusp, and within the temperature window $T_u<T<T(k_s^{crit})$, we have two phases. The red phase, with $k_s^{crit}<k<k_s^{u}$ has always higher free energy for a given temperature. So it is never the thermodynamically preferred phase in the canonical ensemble. Note also that the solution is regular at the cusp (in particular, in the equivalent plot of free energy versus $k_s$, this cusp corresponds to a maximum inflection point of the curve).  
 
Plotting the dimensionless entropy density  $ \frac{4\pi  G_5}{P^2g_0^2}\frac{s}{T^3}$ as a function of the dimensionless energy density $\frac{32\pi^4}{81}\frac{\cal E}{\Lambda^4}$ we find that the two branches of solution can never have the same energy and entropy densities (see {\it Left Panel} of Fig. \ref{figure2}).

\begin{figure}[th]
\centering
\includegraphics[width=.48\textwidth]{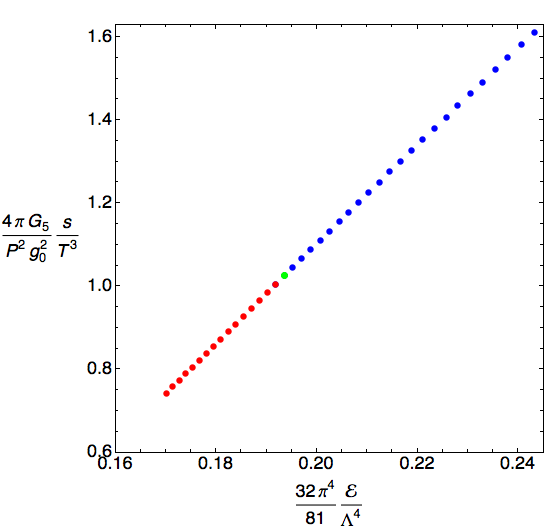}
\hspace{0.5cm}
\includegraphics[width=.46\textwidth]{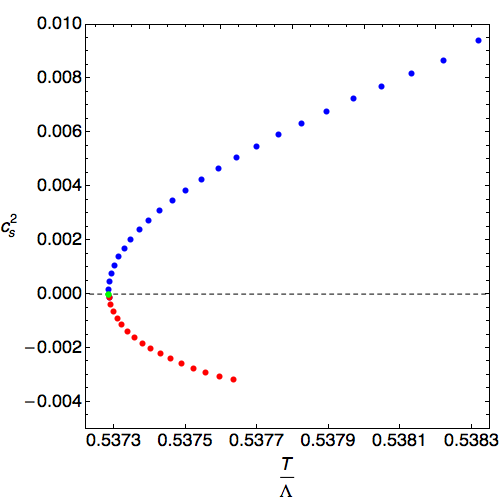}
  \caption{
{\bf Left Panel:}   The dimensionless entropy density as a function of the dimensionless energy density in the vicinity of $T=T_u$.
  {\bf Right Panel:}  Square of speed of sound $c_s^2$ in the vicinity of $T=T_u$.  } \label{figure3}
\end{figure}

So, why is $T=T_u$ so special? The answer relies on the study of speed of sound waves in the cascading plasma \cite{Buchel:2009bh}. In the gravitational description, this speed of sound $c_s$ can be computed from the lowest quasinormal mode dispersion relation in the sound channel. Alternatively, in the dual gauge theory description,
it is computed from the standard thermodynamic relation $c_s=\sqrt{\frac{\partial {\cal P}}{\partial {\cal E}}}$. It turns out to be
 \begin{equation}
c_s^2=\frac 13\ \frac{7-12\,\frac{ a_{2,0}}{a_0}
-6 P^2\, \frac{d a_{2,0}}{d k_s}}{7+4\,\frac{ a_{2,0}}{a_0}+2 P^2\, \frac{da_{2,0}}{d k_s}}  .
\end{equation}
In the vicinity of $T_u\left(k_s^{u}\right)$ (determined by the condition $c_s=0$) we find that $c_s \sim\left( k_s - k_s^{u} \right)$ and  $c_s \sim \left( 1-\frac{T_u}{T}\right)^{1/4}$. Thus, the speed of sound becomes imaginary for $k_s < k_s^{u}$ (red dots), as shown in the {\it right panel}  of  Figure \ref{figure3}. Consequently, the specific heat $c_V=\frac{s}{c_s^2}$ of the cascading plasma diverges at the threshold of the linear instability, $T=T_u$. We conclude that the KT deconfined plasma phase becomes unstable at this temperature. Note that from the inset plots of Fig.  \ref{figure2} one has that $\frac{\partial {\cal F} }{\partial T}{\bigl |}_{T\to T_u}$ is finite while  $\frac{\partial {\cal E} }{\partial T}{\bigl |}_{T\to T_u}$ diverges. This divergency in the change of the energy density is responsible for the vanishing of the speed of sound.  

To summarize, the KT black hole is dual to the chirally-symmetric deconfined phase of the cascading gauge theory. This is the thermodynamically dominant phase of the theory at high temperatures.  As its temperature decreases, we reach a critical point where its pressure (free energy) first vanishes and then becomes negative (positive). At this critical temperature \eqref{confT}, a first order deconfinement/confinement phase transition occurs. This is a non-perturbative phase transition since it proceeds via the nucleation of bubbles of the confined phase. Below this temperature, the KT deconfined chiral symmetric phase is still a metastable phase of the system all the way down to $T_u=0.8749(0) T_c$, where it  joins a perturbatively unstable branch (red dots in the plots) of the theory with negative specific heat.

There is another property of the black hole that is particularly important for our study. 
The KT black hole is also characterized by its D5-brane and D3-brane charges.
The (Maxwell) D5-brane charge, also called the fractional D3-brane charge, is simply
\begin{equation}
Q_{D5}^{Max}=M=\frac{2^{3/4} \pi}{3 G_5^{1/4} } P,
\label{QD5}
\end{equation}
where we used \eqref{defPvsM}.
The definition of the D3-brane charge is more subtle since the logarithm dependence of the fields makes a naive definition diverge, and we need a suitable regularization procedure. 
For a supergravity solution with non-trivial Wess-Zumino terms, one can  define three different types of charges~\cite{Marolf:2000cb, Aharony:2009fc}, that we now discuss. For a solution asymptoting to the KT background and with a horizon, the dimensionless D3-Page charge is ($\alpha^\prime\equiv 1$)\cite{Bena:2011wh}
\begin{equation}\label{defQpage}
Q_{D3}^{Page}= \frac{1}{(4\, \pi^2)^2} \int _{T^{1,1}} \left( {\cal F}_5-B_2\wedge F_3\right),
\end{equation}
where the integration is done over the $T^{1,1}$. The D3-Page charge  is conserved and is independent of the radius at which it is evaluated. 

We can also define the Maxwell D3-charge of the solution as 
\begin{equation}
Q_{D3}^{Max}= \frac{1}{(4\pi^2)^2} \int_{T^{1,1}_{x_c}} {\cal F}_5 \, ,
\end{equation}
where the integral is again performed over  $T^{1,1}$, but this time  with the integrand evaluated at a certain cut-off $x=x_c$.
There are two physically distinct contributions to the Maxwell charge, from the black hole (or from mobile branes when no horizon is present), $Q_b$, and from charge dissolved in the fluxes ($Q_f$):
\bea\label{def:qb}
Q_{D3}^{Max}&=& Q_b^{D3}+Q_f^{D3}\,,  \nonumber \\
Q_b^{D3}&=&\frac{1}{(4\, \pi^2)^2}\, \int_{T^{1,1}_{H}} F_5 \,, \\
Q_f^{D3}&=&\frac{1}{(4\, \pi^2)^2}\, \left( \int_{T^{1,1}_{x_c}}F_5- \int_{T^{1,1}_{H}} F_5\right) =\frac{1}{(4\, \pi^2)^2}\, \int_{{\cal M}_6} H_3\wedge F_3  \,.\nonumber
\eea
where $\int_{T^{1,1}_{H}}$ means that the integrand is evaluated at the horizon $x=1$, and ${\cal M}_6$ is the ``bulk" spacetime spanned by the coordinates on $T^{1,1}$ plus the radial coordinate.
The Maxwell charge depends on the scale at which it is measured, but if we fix a holographic screen, we expect physical processes to preserve its value at the screen. In particular, for a given scale, it must be the same if two solutions are to describe different vacua of the same theory. 

\begin{figure}[th]
\centering
\includegraphics[width=.50\textwidth]{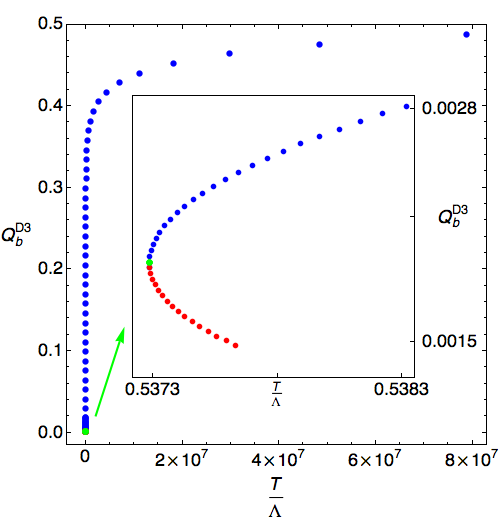}
  \caption{
 The dimensionless D3-brane charge of the black hole $Q_b^{D3}$ as a function of the dimensionless temperature.}  \label{figure4}
\end{figure}

For the KT black hole, we compute the Maxwell charge at the black hole horizon using \eqref{seriesH}, to obtain:
\begin{equation}\label{QD3:KTbh}
Q_b^{D3}=\frac{1}{27\pi }\, k_0^h\,.
\end{equation}
As described previously, the value of $k_0^h$ is determined numerically and it is always positive, for all values of the single parameter $k_s$ that parametrizes the KT black hole family. Therefore, the KT black hole can only have positive Maxwell D3-brane charge at the horizon, as explicitly shown in Fig. \ref{figure4}. In particular, the  D3 charge at the horizon $Q_b^{D3}$ can never have opposite sign to the asymptotic D3 charge $Q_{D3}^{Max}$.

\section{The mass-deformed Klebanov-Strassler black hole\label{sec:KSbh}}

\subsection{Broken-chiral-symmetry phases of the theory. Mass-deformed cascading theories. \label{sec:KSbhPhases}}

So far we have discussed just deconfined/confined phases of the
cascading gauge theory with unbroken chiral symmetry.  From the
holographic gauge theory perspective, the chiral symmetry preservation
is explicitly manifest in the absence of expectation values for
dimension-3 operators in the dual thermal states.  On the other hand,
one knows that at zero temperature the cascading gauge theory confines
into a chiral-symmetry-breaking ($\chi SB$) phase, with the
supergravity dual of this phase being the Klebanov-Strassler (KS)
warped deformed conifold solution \cite{ks}. In the UV region, this
solution asymptotes to the KT solution but, deep into the IR bulk, the
$Z_{2M}$ chiral symmetry is spontaneously broken to $Z_2$ (by gaugino
condensation in the pure SYM limit of the theory) and the KS solution
also breaks the $U(1)_B$ symmetry, while preserving the $SU(2)\times
SU(2)$ symmetry. If we heat the theory, the confined phase persists:
we have now a thermal gas of hadrons described by the KS solution with
a thermal identification. A question that immediately emerges is then
whether this theory also allows for the existence of a
finite-temperature deconfined phase, this time with broken chiral
symmetry, that we would naturally call a KS black hole. This question
does not come alone. {\it \`A priori} there is no reason for this
confinement/deconfinement phase transition of the broken chiral
symmetry phases to occur at the same critical temperature $T_c$ $-$
see \eqref{confT} $-$ of the phase transition for the chiral symmetric
phases.

In supersymmetric gauge theories chiral symmetry breaking necessarily
occurs at temperature at least as high as the confinement temperature.
The question remains concerning the thermal competition of the two
deconfined phases: what is the critical temperature, $T_{\chi SB} $,
at which the KT black hole would exchanges dominance in the partition
function with the KS black hole? That is, what would be the relation
between the critical temperatures for the confinement/deconfinement
and the broken- /unbroken-chiral-symmetry phase transitions?

These questions were analyzed in detail in \cite{Buchel:2010wp}, whose conclusions we summarize next. We will soon realize that some of the above-described intuition needs to be re-evaluated. Consider first the KT deconfined phase with unbroken chiral symmetry. A good way to learn whether this phase can condensate into a broken chiral symmetric phase is to study chiral-symmetry-breaking ($\chi \rm {SB}$) deformations of this phase. In the supergravity description this amounts to study linearized gravitational perturbations about the KT black hole solution that break the chiral symmetry of the background. To make our life easier we want to restrict to the simplest sector of perturbations  that break the desired symmetries while keeping frozen the fluctuations of fields irrelevant for the discussion. Ref.  \cite{Buchel:2010wp} found this sector of deformations that solves the linearized equations of motion of the system. The boundary conditions further constrain the quasinormal mode spectrum of frequencies.  The key observation is that the imaginary part of this frequency spectrum changes sign at a critical temperature   
\begin{equation}
T_{\chi \rm {SB}}=0.882503(0) T_c >T_u\,.
\end{equation} 
Actually this $\chi \rm {SB}$ instability is a Gregory-Laflamme type of instability since it requires the breaking of the translational invariance along the planar spatial directions  transverse to $T^{1,1}$. \footnote{As an interesting side note, recall that the deconfined KT plasma is thermodynamically stable down to $T_u$ and the fact that the $\chi \rm {SB}$ Gregory-Laflamme instability occurs at $T_{\chi \rm {SB}} >T_u$ provides an explicit example of a violation of the Gubser-Mitra conjecture.}
We seem to have all the ingredients to expect a new branch of black holes in a phase diagram of solutions of the theory. That is, at the threshold of $\chi \rm {SB}$ instability,  we expect a natural merger of the KT black hole with what would be the KS black hole.  This is where the above intuition proves to be dramatically wrong. Indeed, Ref.  \cite{Buchel:2010wp} numerically searched for gravitational solutions describing the homogeneous and isotropic states of the cascading plasma with spontaneously broken chiral symmetry. The idea was to construct the KS black hole  by deforming the chirally symmetric KT black hole for $T<T_{\chi \rm{SB}}$ along the tachyonic directions revealed in the linearized analysis. This attempt was   unsuccessful: there are {\it no} KS black hole solutions in the cascading gauge theory.

At this point, we still need to add a twist to the story. So far, we have been silently assuming that the KS deconfined phase we were searching for was  breaking the chiral symmetry {\it spontaneously}. If we are less demanding, we can still search for thermal deconfined phases of the {\it mass-deformed} cascading gauge theory that {\it explicitly} break the chiral symmetry.
To get the mass-deformed cascading gauge theory one introduces the mass terms 
\begin{equation}
\mu_i\equiv \frac{m_i}{\Lambda}\,,\qquad i=1,2\,,
\label{masses}
\end{equation}
for the gauginos (the ${\cal N}=1$ fermionic superpartners of $SU(N+M)\times SU(N)$ gauge bosons). These mass terms explicitly break both the supersymmetry  and the chiral symmetry, and the theory does have a  homogeneous and isotropic deconfined thermal phase. The mass terms \eqref{masses} are the couplings of the two dimension-3 operators that explicitly break the chiral symmetry
\begin{equation}
{\cal O}_3^{j}={\cal O}_3^{j}(\mu_i)\,,\qquad j=1,2\,,
\label{vevs3}
\end{equation}
and, in the chiral limit $\mu_i\to 0$, the expectation values for the condensates vanish as well:
\begin{equation}
\lim_{\mu_i\to 0}\ {\cal O}_3^{j}(\mu_i)=0\,.
\label{chirallimi}
\end{equation}  
In the supergravity description,  the homogeneous and isotropic deconfined broken chiral phase is described by what we call the mass-deformed KS black hole. This black holes was numerically constructed at the  full non-linear level in \cite{Buchel:2010wp}. It is reassuring that the same numerical code that does not find the KS black hole dual to what would be a spontaneously broken chiral symmetric deconfined phase of the theory ($\mu_i= 0$), does find the mass-deformed KS black hole ($\mu_i\neq 0$). This definitely guarantees that the statement that KS black holes do not exist is not a consequence of a somehow incomplete numerical search.

We can now turn back to the linearized study of the $\chi \rm {SB}$ physical excitations of the KT black hole. The original study of  \cite{Buchel:2010wp} included linearized perturbations with the mass-deformations turned-on (these mass-deformations  are dimension-3 operators and can be read from the coefficients of the non-normalizable modes that decay asymptotically as $1/r$; their normalizable partners 
 give the VEVs for these dimension-3 operators \eqref{vevs3}). This study finds that at $T=T_{\chi \rm{SB}}$ the KT black hole becomes linearly unstable. As described below, the threshold of this instability signals a branch-off to a mass-deformed KS black hole in a phase diagram of solutions of the cascading theory. However, if  $\mu_i= 0$, the  KT black hole is still unstable but there is no associated KS black hole. The chiral limit \eqref{chirallimi} suggests that in this black hole background the ``chiral tachyons" condensate with finite momentum: the resulting merging phase (and eventual endpoint state in a time evolution) cannot be a homogeneous and isotropic geometry described by what would be the KS black hole.  

\subsection{The mass-deformed Klebanov-Strassler black hole\label{sec:KSbhConstruction}}

In this section we will briefly review the numerical construction of
the mass-deformed KS black hole \cite{Buchel:2010wp}. This black hole
describes the high-temperature phase of the mass-deformed KS
theory. The zero temperature phase of this theory is dual to the
Kuperstein-Sonnenschein perturbative solution \cite{Kuperstein:2003yt} when the
gaugino masses are equal and to the more general perturbative solution of
\cite{Bena:2011wh, Dymarsky:2011pm} when the masses are arbitrary. 
The infrared expansion of the zero-temperature solution is included in the class of perturbative IR solutions constructed in \cite{Bena:2012bk,McGuirk:2009xx}. Our
purpose is to calculate the D3-brane charge at the horizon of the
black hole in this theory and to see whether the sign of the black
hole charge could somehow be made negative by scanning through the
possible ranges of gaugino masses.

We are interested in the most general field ansatz of the cascading gauge theory that describes its homogeneous and isotropic states, both at zero and non-zero temperature. Such ansatz is tailored to search for the (mass-deformed) KS black holes and naturally describes also the supersymmetric Klebanov-Strassler (KS) warped deformed conifold solution \cite{ks}. It also includes as special cases the KT black hole solution and the KT singular supersymmetric solution discussed in the previous section, which further obey the constraints imposed by requiring 
the  $Z_{2M}$ chiral and $U(1)_B$ symmetries.
In the Einstein-frame this ansatz is 
\begin{equation}
ds_{10}^2 =g_{\mu\nu}(y) dy^{\mu}dy^{\nu}+ds^2_{T^{1,1}}\,, \qquad ds^2_{T^{1,1}}=\om_1^2(y) g_5^2
+\om_2^2(y) \left[g_3^2+g_4^2\right]+\om_3^2(y) \left[g_1^2+g_2^2\right],
\label{10metks}
\end{equation}
for the gravitational field ($y$ denotes the coordinates of $\calm_5$ with greek indices $\mu,\nu=0,\cdots,4$), and 
\begin{equation}
\begin{split}
&B_2=h_1(y)\ g_1\wedge g_2+h_3(y)\ g_3\wedge g_4\,,\\
&F_3=\frac 19 P\ g_5\wedge g_3\wedge g_4+h_2(y)\ \left(g_1\wedge g_2-g_3\wedge g_4\right)\wedge g_5
+\left(g_1\wedge g_3+g_2\wedge g_4\right)\wedge d\left(h_2(y)\right)\,,\\
&F_5=\calf_5+\star \calf_5\,,\qquad \calf_5= \left[ 4{\om}_0+ h_2(y)\left(h_3(y)-h_1(y)\right)+\frac 19 P h_1(y) \right] g_5 \wedge
g_3\wedge g_4\wedge 
g_1\wedge g_2\,,\\
&\Phi= \Phi(y)\,,
\end{split}
\label{fluxformsKS}
\end{equation}
for the fluxes $H_3\equiv d B_2$, $F_3$,  $F_5$, and dilaton $\Phi$. 
$P$ is again an integer corresponding to the RR 3-form flux on the compact 3-cycle
(and to the number of fractional branes on the conifold), as given in \eqref{defPvsM}.

To make closer contact with the original study of the KS warped deformed conifold \cite{ks}
we use the 1-forms on $T^{1,1}$,
\begin{equation}
g_1=\frac{\alpha^1-\alpha^3}{\sqrt 2}\,,\qquad g_2=\frac{\alpha^2-\alpha^4}{\sqrt 2}\,,\qquad g_3=\frac{\alpha^1+\alpha^3}{\sqrt 2}\,,\qquad g_4=\frac{\alpha^2+\alpha^4}{\sqrt 2}\,,\qquad g_5=\alpha^5\,,
\label{3form1}
\end{equation}
where 
\begin{equation}
\begin{split}
&\alpha^5=d\psi+\cos\theta_1 d\phi_1+\cos\theta_2 d\phi_2\,,\qquad \alpha^1=-\sin\theta_1 d\phi_1\,,\qquad \alpha^2=d\theta_1\,,\\
&\alpha^3=\cos\psi\sin\theta_2 d\phi_2-\sin\psi d\theta_2\,,\qquad \alpha^4=\sin\psi\sin\theta_2 d\phi_2+\cos\psi d\theta_2\,.
\end{split}
\label{3form2}
\end{equation}
instead of those in \eqref{1forms}.

With this ansatz, a Kaluza-Klein reduction of the type IIB action to five dimensions yields the effective action \cite{Buchel:2010wp}, 
\begin{equation}
\begin{split}
S_5=& \frac{108}{16\pi G_5} \int_{\calm_5} {\rm vol}_{\calm_5}\ \Omega_1 \Omega_2^2\Omega_3^2\ 
\biggl\lbrace 
 R_{10}-\frac 12 \left(\nabla \Phi\right)^2\\
&-\frac 12 e^{-\Phi}\left(\frac{(h_1-h_3)^2}{2\Omega_1^2\Omega_2^2\Omega_3^2}+\frac{1}{\Omega_3^4}\left(\nabla h_1\right)^2
+\frac{1}{\Omega_2^4}\left(\nabla h_3\right)^2\right)
\\
&-\frac 12 e^{\Phi}\left(\frac{2}{\Omega_2^2\Omega_3^2}\left(\nabla h_2\right)^2
+\frac{1}{\Omega_1^2\Omega_2^4}\left(h_2-\frac P9\right)^2
+\frac{1}{\Omega_1^2\Omega_3^4} h_2^2\right)
\\
&-\frac {1}{2\Omega_1^2\Omega_2^4\Omega_3^4}\left(4{\Omega}_0+ h_2\left(h_3-h_1\right)+\frac 19 P h_1\right)^2
\biggr\rbrace\,,\\
\end{split}
\label{KS:5action}
\end{equation}
where $R_{10}$ is given by in terms of the 5-dimensional Ricci scalar of the metric $g_{\mu\nu}$ as
\begin{equation}
\begin{split}
R_{10}=R_5&+\left(\frac{1}{2\Omega_1^2}+\frac{2}{\Omega_2^2}+\frac{2}{\Omega_3^2}-\frac{\Omega_2^2}{4\Omega_1^2\Omega_3^2}
-\frac{\Omega_3^2}{4\Omega_1^2\Omega_2^2}-\frac{\Omega_1^2}{\Omega_2^2\Omega_3^2}\right)-2\Box_5 \ln\left(\Omega_1\Omega_2^2\Omega_3^2\right)\\
&-\biggl\{\left(\nabla\ln\Omega_1\right)^2+2\left(\nabla\ln\Omega_2\right)^2
+2\left(\nabla\ln\Omega_3\right)^2+\left(\nabla\ln\left(\Omega_1\Omega_2^2\Omega_3^2\right)\right)^2\biggr\}\,,
\end{split}
\label{KS:ric5}
\end{equation}
and $\nabla$ and $\square_{\,5}$ denote the covariant derivative and the Laplacian  in $\calm_5$, respectively.
This action \eqref{KS:5action} also describes the theory governed by \eqref{5action} as a special limit \cite{Buchel:2010wp}.

The general 5-dimensional background geometry with homogeneous and isotropic (but not necessary
Lorentz-invariant) asymptotic boundary takes the form written in \eqref{ktm}. It is tailored to search for the (mass-deformed) KS black hole with horizon located at $x=1$ and asymptotic KT background at $x\to 0$.
It is also useful to encode all the information convening the warp-squashed $T^{1,1}$ factors $\Omega_{1,2,3}(x)$,  the flux-form functions $h_{1,2,3}(x)$ and  the dilaton $\Phi(x)$ in the following set of new functions  $\{K_1,\,K_2,\, K_3,\, f_a,\, f_b,\, f_c,\, h,\, g\}$ defined as:
\begin{eqnarray}
&&\Omega_i=\omega_i H^{1/4}\quad (i=1,2,3),
\nonumber \\
&& \hspace{1cm} H(x)=(2x-x^2)\ h(x)\,, \nonumber \\
&& \hspace{1cm}  \omega_1(x)=\frac{\sqrt{f_c(x)}}{3(2x-x^2)^{1/4}}\,,\qquad \omega_2(x)=\frac{\sqrt{f_a(x)}}{\sqrt{6}(2x-x^2)^{1/4}}\,,\qquad \omega_3(x)=\frac{\sqrt{f_b(x)}}{\sqrt{6}(2x-x^2)^{1/4}}\,; \nonumber \\
&&h_1(x)=\frac 1P \left(\frac{1}{12}\ K_1(x)-36 \Omega_0\right)\,,\quad h_2(x)=\frac{P}{18}\ K_2(x)\,, 
\quad h_3(x)=\frac 1P \left(\frac{1}{12}\ K_3(x)-36 \Omega_0\right);\nonumber \\
&& g(x)=e^{\Phi(x)}\,.
\label{deff2}
\end{eqnarray}

From the equations of motion of \eqref{KS:5action} we find that $G_{xx}(x)$ can be expressed by an algebraic relation that is a function of $\{K_1,\,K_2,\, K_3,\, f_a,\, f_b,\, f_c,\, h,\, g\}$  and their first derivatives. The non-trivial equations of motion can then be written as a system of eight second order non-linear coupled ODEs for $\{K_1,\,K_2,\, K_3,\, f_a,\, f_b,\, f_c,\, h,\, g\}$. Each of these eight equations is second order. It follows that the total differential order of the system is $16$. The constant $\Omega_0$, introduced in  \eqref{fluxformsKS} does not appear in the equations of motion.

The (mass-deformed) KS black hole must solve these equations of motion subject to appropriate boundary conditions at the horizon, $x\to 1$, and at the asymptotic boundary $x\to 0$.

Near the horizon ($x\to 1$), the equations of motion have the IR series expansion:
\begin{equation}
\begin{split}
&K_i=\sum_{n=0}^\infty k_{in}^h\ (1-x)^{2n}\,,\qquad i=1,2,3\,,\\
&f_{\alpha}=a_0\ \sum_{n=0}^\infty f_{\alpha n}^h\ (1-x)^{2n}\,,\qquad \alpha=a,b,c\,,\\
&h=\sum_{n=0}^\infty h_{n}^h\ (1-x)^{2n}\,,\qquad g=g_0\ \sum_{n=0}^\infty g_{n}^h\ (1-x)^{2n}\,.
\end{split}
\label{seriesH:KS}
\end{equation}
and regularity of the expansion in the ingoing Eddington-Finkelstein coordinates requires the leading terms of this expansion to be just regular. There are 9 independent IR parameters that we choose to be: 
\begin{equation}
\hbox{IR independent parameters (9):}\qquad   \{k_{1h0}\,, k_{2h0}\,, k_{3h0}\,,f_{ah0}\,, f_{ah1}\,, f_{bh0}\,, f_{ch0}\,, h_{h0}\,, g_{h0}\}\,.
\label{KS:paramH}
\end{equation}

At the UV asymptotic boundary we want the (mass-deformed) KS black hole to asymptote to the KT solution \cite{kt}, while allowing also for the mass-deformation parameters. The latter is the leading order contribution of a power series expansion in $x$ and $\ln(x)$ of the equations of motion:
\begin{equation}
\begin{split}
K_1=& 4 h_0a_0^2-\frac 12 P^2 g_0-\frac12 P^2 g_0\ \ln x+\sum_{n=1}^\infty\sum_k\ k_{1nk}\ x^{n/4}\ \ln^k x\,, \\
K_2=& 1+\sum_{n=1}^\infty\sum_k\ k_{2nk}\ x^{n/4}\ \ln^k x\,,\\
K_3=& 4 h_0a_0^2-\frac 12 P^2 g_0-\frac12 P^2 g_0\ \ln x+\sum_{n=1}^\infty\sum_k\ k_{3nk}\ x^{n/4}\ \ln^k x\,,\\
f_\alpha=& a_0\biggl(1+\sum_{n=1}^\infty\sum_k\ f_{\alpha nk}\ x^{n/4}\ \ln^k x\biggr)\,,\qquad\alpha=\{a,\,b\}\\
f_c=& a_0\biggl(1+\sum_{n=2}^\infty\sum_k\ f_{cnk}\ x^{n/4}\ \ln^k x\biggr)\,,\\
h=& h_0-\frac  {P^2 g_0}{8a_0^2}\ \ln x+\sum_{n=2}^\infty\sum_k\ h_{nk}\ x^{n/4}\ \ln^k x\,,
\qquad  g= g_0\left( 1+\sum_{n=2}^\infty\sum_k\ g_{nk}\ x^{n/2}\ \ln^k x \right)\,. 
\end{split}
\label{KS:seriesBdry}
\end{equation}
There are 6 independent microscopic parameters. Four of them describe the KT solution, namely $\{P,\,g_0,\,a_0,\,h_0 \}$, and have exactly the physical interpretation already described below \eqref{KS:seriesBdry}. In particular, $h_0$ can be traded for the dimensionless parameter $k_s$, defined in \eqref{def:ks}, which defines  the dynamical scale $\Lambda$ of the cascading theory. Moreover, the straightforward extension (to accommodate for the field content of the KS ansatz)  of the three scaling symmetries discussed in \eqref{scaling}-\eqref{scaling2} still allows to set \begin{equation}
P=g_0=a_0\equiv 1\,.
\label{reduce}
\end{equation}
The other 2 independent microscopic parameters are $\{ k_{110},\,f_{a10}\}$ and are related to  the couplings of the two dimension-3 operators that explicitly break the chiral symmetry. The explicit relation between $\{ k_{110},\,f_{a10}\}$ and these two mass-deformation parameters $\{ \mu_1,\,\mu_2 \}$ of the cascading gauge theory introduced in \eqref{masses}  is 
\begin{equation}
f_{a10}=\left(\mu_1+4 \mu_2\ k_s\right)\ e^{-k_s/2}\,,\qquad \left(k_{110}+\frac 12 P^2 g_0 f_{a10}\right)\left(3P^2 g_0+8 h_0 a_0^2\right)^{-1}=\mu_2\ e^{-k_s/2}.
\label{kf:mu}
\end{equation}
In addition to these microscopic UV parameters, there are 7 extra independent parameters $\{f_{a30}\,, k_{230}\,, f_{a40}\,, g_{40}\,, f_{a60}\,, f_{a70}\,, f_{a80}\}$ associated to the VEV's of the operators dual to the fields we are solving for. 
Summarizing, after using the scaling symmetries of our system, we are left with 10 UV independent parameters,
\begin{equation}
\hbox{UV independent parameters (10):}\qquad  \{\,k_s,\,k_{110},\,f_{a10},\,f_{a30}\,, k_{230}\,, f_{a40}\,, g_{40}\,, f_{a60}\,, f_{a70}\,, f_{a80}\}.
\label{KS:paramBdry}
\end{equation}

The total number of IR  and UV independent parameters is $9+10=19$ and is given by \eqref{KS:paramH} and \eqref{KS:paramBdry}. On the other hand the equations of motion are a system of total differential order 16. 
Therefore, a general mass-deformed KS black hole is a 3-parameter family of solutions. We naturally take these 3 parameters to be the microscopic parameters of the cascading gauge theory, namely $\{ k_s,\,k_{110},\,f_{a10}\}$. We can then use  \eqref{def:ks} and \eqref{kf:mu} to express the results in terms of the temperature  $T/\Lambda$ and the mass-deformation parameters $\{ \mu_1,\,\mu_2 \}$. Of course, if we are just interested in the KS black hole that would describe the spontaneously symmetry broken deconfined phase, we set $\mu_1=\mu_2=0$ and we have simply a 1-parameter family of solutions. As emphasized previously, the numerical code does find generic mass-deformed KS black holes (whose properties will be discussed below), but when we crank down the mass-deformations to zero, no solution with $\calo_3^j\ne 0$ is found. 

The numerical construction of the mass-deformed KS black holes is done using a shooting method similar to the one used in the construction of the KT black hole. In this shooting method,
the IR power series solution \eqref{seriesH:KS} is constructed up to order $n=1$ (inclusive),
and the asymptotic UV series solution \eqref{KS:seriesBdry} is constructed up to eighth order ($n=8$, inclusive). 
One then evolves these solutions away from these critical points and do the matching in the overlapping intermediate region (at $x=0.5$). Fixing the microscopic parameters $\{ k_s,\,k_{110},\,f_{a10}\}$, we still have 16 independent IR/UV parameters that are determined by the 16 conditions that follow from requiring that the 8 functions  $\{K_1,\,K_2,\, K_3,\, f_a,\, f_b,\, f_c,\, h,\, g\}$ and their first derivatives are continuous at $x=0.5$.
The process is repeated for different values of $\{ k_s,\,k_{110},\,f_{a10}\}$.

\subsection{Properties of the mass-deformed Klebanov-Strassler black hole\label{sec:KSbhProperties}}

\begin{figure}[t]
\centerline{\includegraphics[width=.50\textwidth]{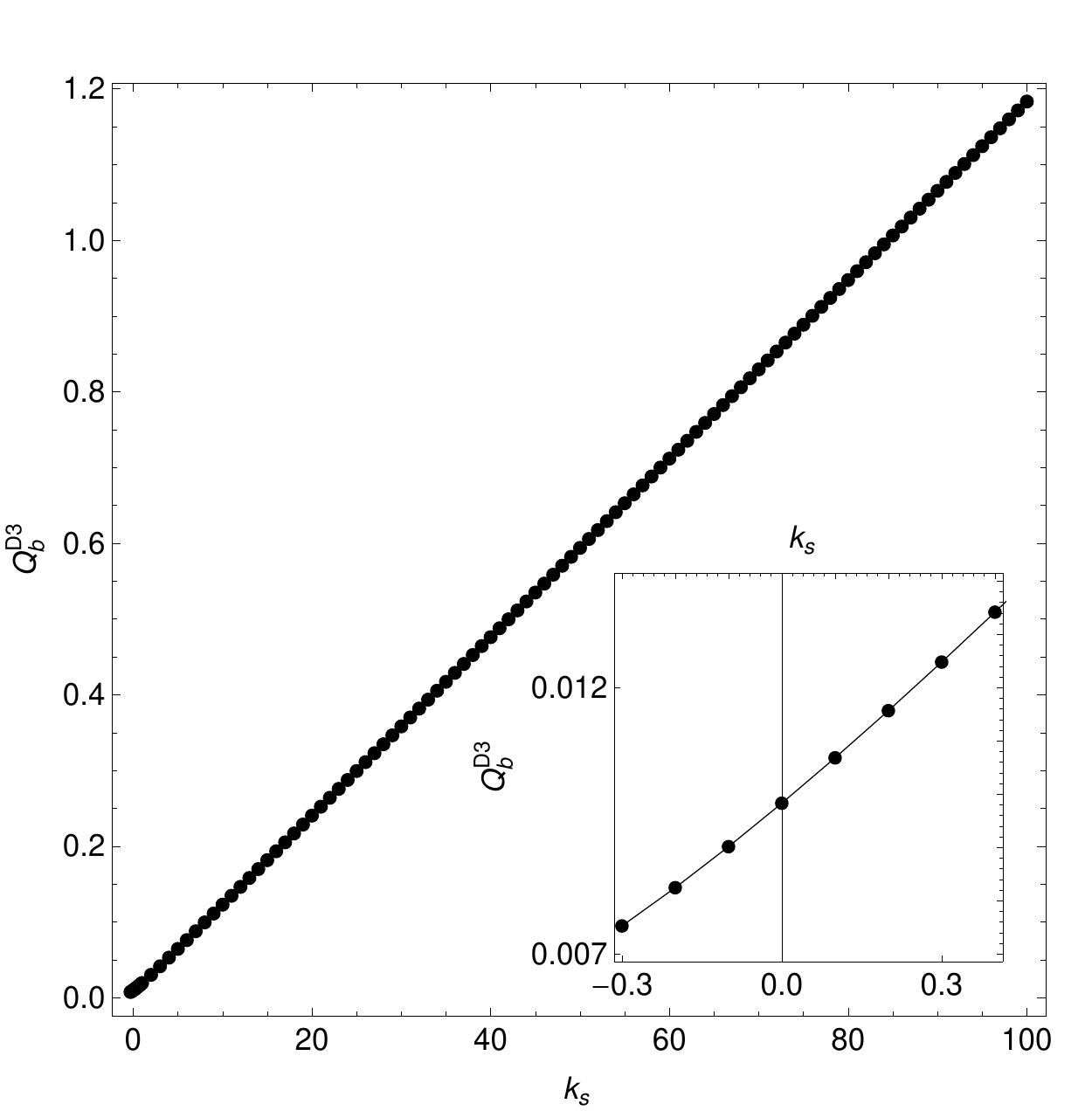}}
\caption{
 Mobile D3-brane charge $Q_b^{D3}$ for mass-deformed KS black holes with mass-deformation parameters  $f_{a10}=0.1$ and $k_{110}=0$, as a function of  $k_s$ (related to $T/\Lambda$).
} \label{fig:KScharge1}
\end{figure}

\begin{figure}[t]
\begin{center}
  \includegraphics[width=3in]{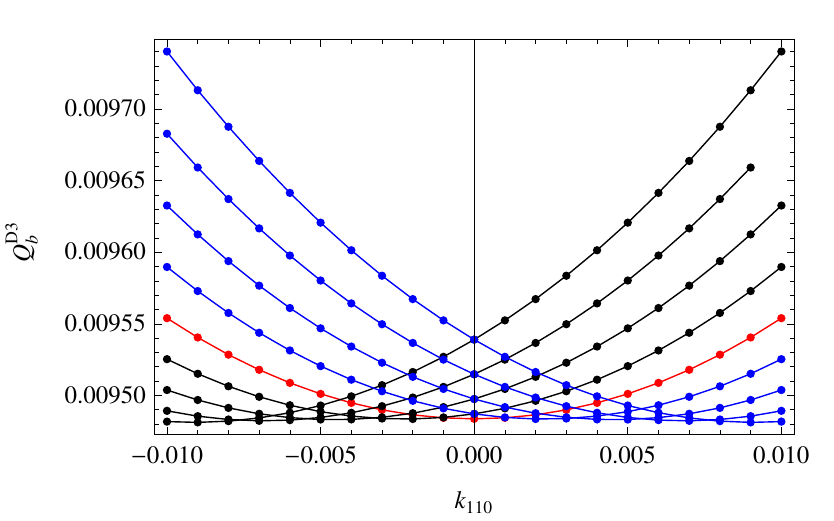}
  \includegraphics[width=3in]{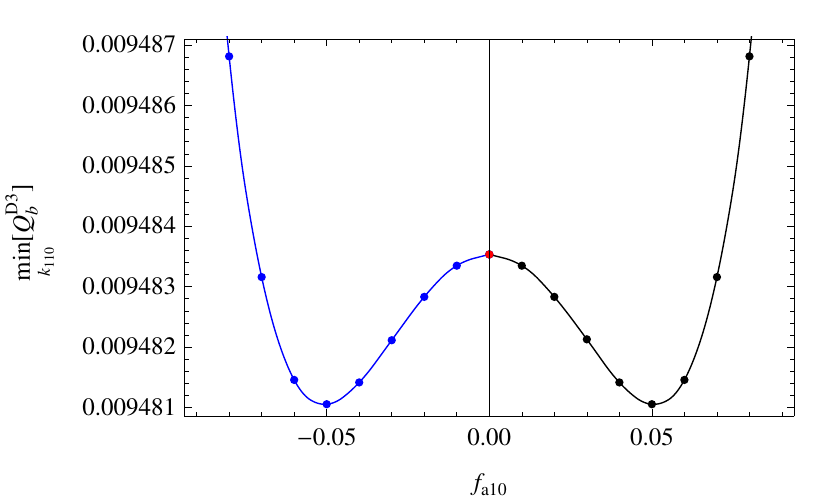}
\end{center}
  \caption{
{\it \bf Left Panel:} Mobile D3-brane charge $Q_b^{D3}$ for
mass-deformed KS black holes with parameters $k_s=0$ and $f_{a10}=\{-0.4,\cdots,0.4\}$, as a function of the
mass-deformation parameter $k_{110}$. The central red curve/dots with
minimum at $k_{110}=0$ is for $f_{a10}=0$. As $f_{a10}$ grows
increasingly positive, the minimum shifts to the left (black curves
are for $f_{a10}=\{0.1,\cdots,0.4\}$). As $f_{a10}$ grows increasingly
negative, the minimum shifts to the right (blue curves are for
$f_{a10}=\{-0.4,\cdots,-0.1\}$).  {\it \bf Right Panel:} Minima of the black hole D3-brane
charge $Q_b^{D3}$ for mass-deformed KS black holes with respect to $k_{110}$ for 
$k_s=0$, as a function of the
mass-deformation parameter $f_{a10}$.
}\label{fig:KScharge2}
\end{figure}

\begin{figure}[t]
\begin{center}
  \includegraphics[width=3in]{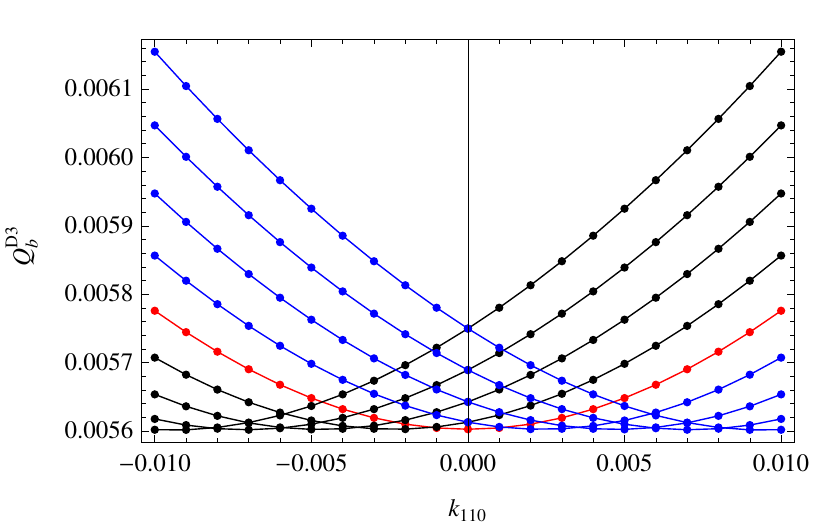}
  \includegraphics[width=3in]{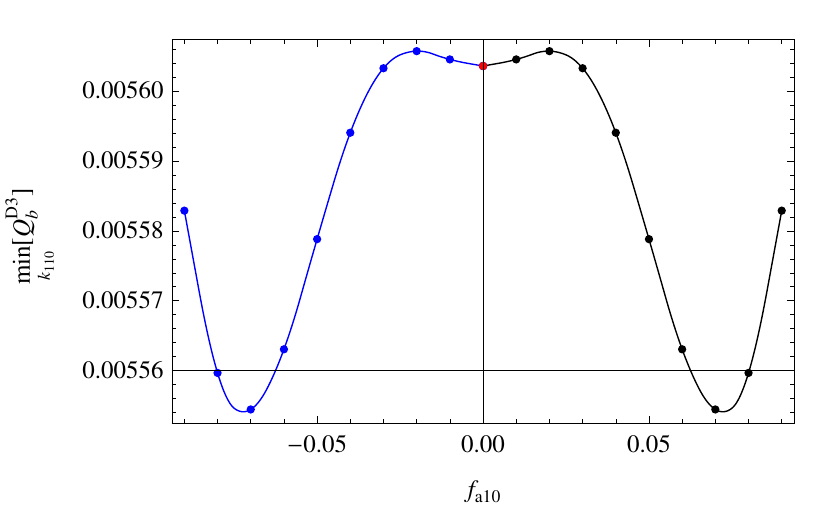}
\end{center}
  \caption{
Same as Fig.~\ref{fig:KScharge2}, but with $k_s=-0.5$.
}\label{fig:KScharge3}
\end{figure}

\begin{figure}[t]
\begin{center}
  \includegraphics[width=3in]{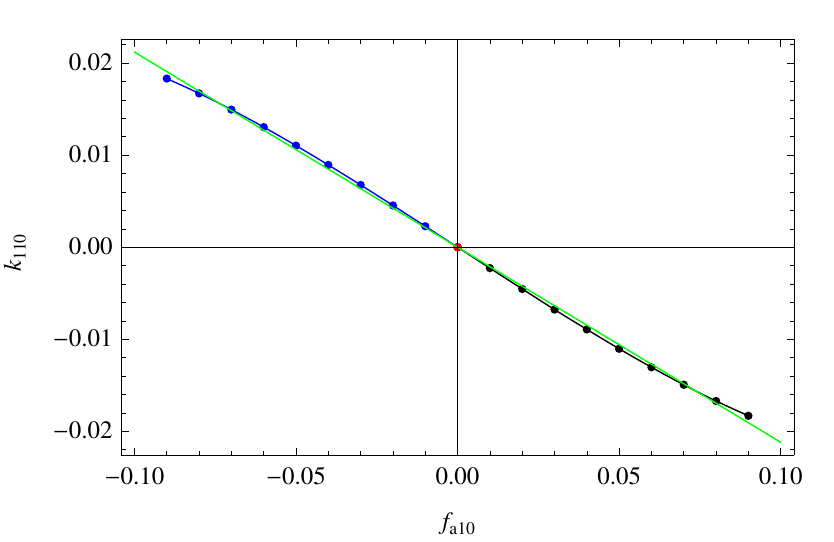}
  \includegraphics[width=3in]{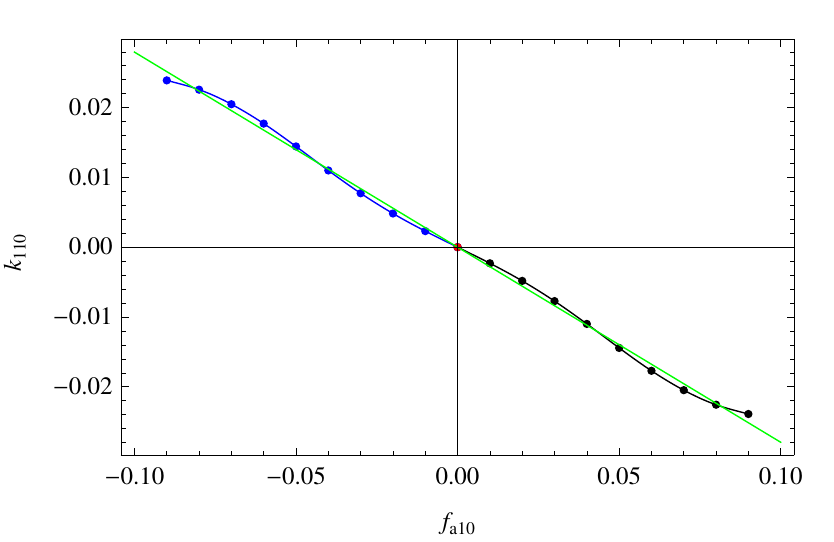}
\end{center}
  \caption{
Relation between the mass-deformation parameters $f_{a10}$ and 
$k_{110}$ for $k_s=0$ ({\bf Left Panel}) and $k_s=-0.5$ 
({\bf Right Panel}) at  $\min Q_b^{D3}$. The solid green lines 
represent the best fit linear relation  $f_{a10}\ \propto k_{110}$.
 }\label{fig:k110fa10}
\end{figure}

Two quantities of utmost interest for our discussion are the  D5-brane and D3-brane charges of the mass-deformed KS black hole. The former is simply given by \eqref{QD5}, $Q_{D5}^{Max}=M$. On the other hand, we can compute the Maxwell D3 charge of the black hole \eqref{def:qb} using \eqref{seriesH:KS} to get
\begin{equation}\label{QD3:KSbh}
Q_b^{D3}=\frac{ k_{10}^h \left(2-k_{20}^h\right)+k_{20}^h \,k_{30}^h}{54 \pi } \,.
\end{equation}
Note that the constant constant $\Omega_0$ introduced in \eqref{fluxformsKS} does not appear in the equations of motion neither in the flux $F_5$ but it does appear in the contribution $B_2\wedge F_3$ to the D3-Page charge \eqref{defQpage}. 
Below we present the properties of the mass-deformed KS black hole solution. As described above this is a 3-parameter family of solutions parametrized by $\{ k_s,\,k_{110},\,f_{a10}\}$, the values of which we give as an input to the code. The values of $\{ k_{10}^h,\,k_{20}^h,\,k_{30}^h\}$ are then determined numerically. For the several families of values of $\{ k_s,\,k_{110},\,f_{a10}\}$ that we have tried, it is always true that $Q_b^{D3}>0$.

We outline now our searches. In Fig.~\ref{fig:KScharge1} we present values of $Q_b^{D3}$ 
with mass-deformation parameters  $f_{a10}=0.1$ and $k_{110}=0$, as a function of  $k_s$ (related to $T/\Lambda$). 
Note that smaller values of $k_s$ bring down the Maxwell D3-brane charge at the horizon --- it is thus desirable 
to study as small values of $k_s$ as possible. The technical reason that restricts this is rooted in
the instability of the finite-difference 
codes\footnote{It is possible that these instabilities can be alleviated with spectral 
methods. We hope to report on this in future work.} used to simulate mass-deformed KS black holes 
for small $k_s$. The insert in Fig.~\ref{fig:KScharge1} shows the smallest  values of $k_s$ 
we can trust with our current numerics. A quadratic fit (in $k_s$) to these 8 points shows that 
\begin{equation}
\min_{k_s}\biggl[ Q_b^{D3}\biggl(f_{a10}=0.1,k_{110}=0;\ k_s\biggr)\biggr]\ =\  0.000323195\,.
\label{minqb1}
\end{equation} 
Figs.~\ref{fig:KScharge2} and \ref{fig:KScharge3} present the results of a 
different search strategy. Namely, we fix $k_s$ (set $k_s=0$ for Fig.~\ref{fig:KScharge2} and 
$k_s=-0.5$ for Fig.~\ref{fig:KScharge3}), and generate families of mass-deformed  
KS black holes by scanning the mass-deformation parameter $k_{110}$ for discrete 
set of $f_{a10}=\{-0.4,\cdots 0.4\}$ (see the {\it{Left panels}}). For each of these curves we use 
quadratic extrapolation to determine $\min_{k_{110}} [Q_b^{D3}]$ (see the {\it{Right panels}}). 
Finally, Fig.~\ref{fig:k110fa10} indicate functional relations between non-normalizable 
modes $f_{a10}$ and $k_{110}$ at minima of $Q_b^{D3}$ for $k_s=0$  (see the {\it{Left panel}})
and $k_s=-0.5$  (see the {\it{Right panel}}). The solid green lines represent the best linear 
fit $k_{110}\propto f_{a10}$. Note that  \eqref{kf:mu} 
implies 
\begin{equation}
k_{110}=f_{a10}\ \frac{4\mu_2-\frac{\mu_1}{2}}{\mu_1+4\mu_2 k_s}\,.
\label{fa10k110resl}
\end{equation}
Thus, a linear relation between $k_{110}$ and $f_{a10}$, if true, would imply a linear 
relation between $m_1$ and $m_2$. From the fits we determine:
\begin{equation}
\frac{m_2}{m_1}= 
\begin{cases}
0.072\,,\ {\rm for}\ k_s=0\,;\cr
0.064\,,\ {\rm for}\ k_s=-0.5\,.
\end{cases}
\label{fitresults}
\end{equation} 
Assuming that discrepancy between $\frac{m_2}{m_1}$ for different $k_s$ is a numerical artifact, 
an interesting question remains as to physical significance of this 
ratio.

Although we cannot exclude the possibility that there is some region
in the parameter space that yields $Q_b^{D3}<0$, we could not find it.
We take this, together with the fact that the KT black hole of the
previous section always has positive $Q_b^{D3}$ (recall that for this solution
we could run the full parameter space since it is a one-parameter
solution), as strong evidence for the non-existence of KS black holes
with $Q_b^{D3}<0$. To summarize, the mass-deformed KS black holes we have examined always have
positive Maxwell D3-brane charge at the horizon; there appear to be no such black
holes that have a Maxwell charge whose sign is opposite to that of the asymptotic charge of the solution. 
\section{Conclusion\label{conclusion}}

The idea of a Landscape of deSitter vacua in String Theory is founded on two pillars. The first is the construction of a very large number of string theory compactifications with stabilized moduli, either via the KKLT construction \cite{Kachru:2003aw} or by other mechanism \cite{Balasubramanian:2005zx}, which always give anti-deSitter vacua. The second is the uplift of these vacua to deSitter, and the most common way to do this is to trap anti-D3-branes in warped-deformed conifold-like regions of the compactification manifold \cite{Kachru:2003aw}. The recent analysis of \cite{Bena:2009xk,Bena:2011hz,Bena:2011wh,Bena:2012bk} shows that the geometries corresponding to these antibranes are singular, and that furthermore this singularity does not appeared to be resolved \`a la Polchinski-Strassler \cite{Polchinski:2000uf}, which is its most obvious resolution channel \cite{saclay-recent}; this supports the idea that this singularity is not physical, and should therefore be discarded. 

In this paper we have looked at the issue from a different 
perspective: If the singularities coming from anti-D3-branes
at the bottom of the Klebanov-Strassler throat are physical, then, 
following \cite{Gubser:2000nd}, it should be possible to 
cloak them with a regular Schwarzschild horizon. The resulting 
nonextremal geometries would be Klebanov-Tseytlin 
or Klebanov-Strassler black holes  that have negative D3 brane 
Maxwell charge $Q_b^{D3}$ at the horizon. These black holes were 
constructed numerically in \cite{Aharony:2007vg,Buchel:2009bh,Buchel:2010wp} and 
we did an extensive scan of all these numerical solutions but were unable to find any solution with 
$Q_b^{D3}<0$. While not a rigorous proof, our negative result strongly supports the idea that singularities generated by anti-D3-branes at the bottom of the warped deformed conifold 
are unphysical. This in turn suggests that the most common mechanism for uplifting the landscape of AdS vacua one obtains from string theory compactifications with stabilized moduli to deSitter does not work, and hence string theory may only have a landscape of AdS vacua but not a landscape of deSitter vacua.

Furthermore, we also found that the charge of a given KT or mass-deformed-KS black hole 
is not an independent parameter, but is completely determined by the temperature and the gaugino masses. 
This implies that if one is to perform a gedanken experiment that consists of lowering an anti-D3 brane 
into this black hole keeping the temperature fixed, the charge of this black hole will become a bit smaller for a moment, 
but then will immediately go back to its previous value by absorbing charge from the surrounding fluxes. Hence, this black hole acts as a catalyst for brane-charge annihilation for arbitrarily small values of the temperature, and this may suggest that adding an antibrane to a vacuum Klebanov-Strassler solution will cause the surrounding flux to immediately annihilate it \cite{Blaback:2012nf}.

It would be interesting to further push 
the limits of the KT/KS black hole parameter space
(driven by the development of more efficient 
numerical techniques) in order to establish more completely that regular nonextremal 
geometries with $Q_b^{D3}<0$ do not exist. 
It would also be interesting to 
compute D3 brane charge of the bottom of 
de-Sitter deformed KT/KS geometries  
\cite{Buchel:2002wf,Buchel:2006em}. We hope to report 
on this in future work.

%
%
~\\
\noindent{\bf Acknowledgments:}
It is a pleasure to thank Gregory Giecold, Mariana Gra\~na, Stanislav Kuperstein, Stefano Massai, Al Muller, Jorge Santos and Thomas van Riet
for interesting and helpful discussions. This work was supported in part by the ANR grant 08-JCJC-0001-0, the ERC Starting Grant 240210 - String-QCD-BH and by NSERC Discovery Grants.  Research at Perimeter Institute is supported by the Government of
Canada through Industry Canada and by
the Province of Ontario through the Ministry of Research \& Innovation.
AB and OD would like to thank the Institut de Physique Th\'eorique, CEA-Saclay and the Perimeter Institute respectively for hospitality during various stages of this work.
OD further thanks the Yukawa Institute for Theoretical Physics (YITP) at Kyoto University, where part of this work was completed during the YITP-T-11-08 programme  ``Recent advances in numerical and analytical methods for black hole dynamics", and the  participants of the workshops ``The Holographic Way: String Theory, Gauge Theory and Black Holes" at Nordita, ``Spanish Relativity Meeting in Portugal",  ``Exploring AdS-CFT Dualities in Dynamical Settings" at the Perimeter Institute and ``Holography, Gauge Theory and Black Holes" at IOP Amsterdam for discussions.


\end{document}